# IA aplicada al análisis del conflicto Irán-Israel: Mapeo de discursos en YouTube

AI Applied to the Analysis of the Irán-Israel Conflict: Mapping Discourses on YouTube

**Alvaro Vallejo Ramírez**
(Bachelor's Degree in Education – Independent Researcher)
alvaro.vallejo.04@unsch.edu.pe
ORCID: 0009-0004-0600-3701

**Resumen**

**Propósito.** Representación digital del conflicto Irán–Israel ocurrido en junio de 2025, a partir de 120,000 comentarios publicados en YouTube. Se buscó identificar las posturas discursivas en torno a los actores implicados y examinar cómo los sesgos mediáticos y algorítmicos configuran la conversación digital. **Metodología.** Se adoptó un diseño mixto con triangulación. En la fase cuantitativa se emplearon técnicas de procesamiento de lenguaje natural y modelos de aprendizaje automático (BERT y XLM-RoBERTa) para clasificar los comentarios en diez categorías. En la fase cualitativa se desarrolló un análisis crítico del contexto mediático y de las narrativas ideológicas, complementado con anotación manual y entrenamiento supervisado. Esta estrategia permitió integrar robustez estadística y comprensión contextual. **Resultados y conclusiones**. Los hallazgos muestran una clara sobrerrepresentación de discursos pro-Palestina y anti-Estados Unidos/Israel, mientras que las posturas pro-Estados Unidos y anti-Palestina fueron marginales. Irán, habitualmente invisibilizado en los medios globales, emergió como actor central en la conversación digital durante el conflicto, lo que sugiere un cambio narrativo respecto de marcos hegemónicos previos. Asimismo, los resultados confirman la influencia de sesgos algorítmicos en la amplificación de ciertos discursos y la limitación de otros. **Aportes originales**. Este trabajo combina análisis computacional y crítica filosófica para el estudio de controversias digitales, aportando un marco metodológico replicable en contextos geopolíticos. Es uno de los primeros estudios en español en mapear, mediante inteligencia artificial y análisis crítico, los discursos sobre un conflicto internacional en YouTube, visibilizando asimetrías y disputas narrativas poco atendidas.

**Palabras clave:** conflicto Irán-Israel, discurso digital, guerra de los 12 días, inteligencia artificial, sesgo de YouTube.

**Abstract**

Purpose. This study analyzes the digital representation of the Iran–Israel conflict that occurred in June 2025, based on 120,000 comments posted on YouTube. It sought to identify discursive positions regarding the actors involved and to examine how media and algorithmic biases shape digital conversations. Methodology. A mixed-methods design with triangulation was adopted. In the quantitative phase, natural language processing techniques and machine learning models (BERT and XLM-RoBERTa) were used to classify comments into ten categories. In the qualitative phase, a critical analysis of media context and ideological narratives was conducted, complemented by manual annotation and supervised training. This strategy enabled the integration of statistical robustness with contextual understanding. Results and conclusions. The findings reveal a clear overrepresentation of pro-Palestinian and anti-United States/Israel discourses, while pro-United States and anti-Palestinian positions were marginal. Iran, usually rendered invisible in global media, emerged as a central actor in the digital conversation during the conflict, suggesting a narrative shift away from previous hegemonic frameworks. Likewise, the results confirm the influence of algorithmic biases in amplifying certain discourses while limiting others. Original contributions. This work combines computational analysis and philosophical critique for the study of digital controversies, providing a methodological framework replicable in geopolitical contexts. It is one of the first Spanish-language studies to map, through artificial intelligence and critical analysis, discourses on an international conflict on YouTube, highlighting asymmetries and narrative disputes that are often overlooked.

# INTRODUCCIÓN

El 13 de junio de 2025, en una operación relámpago y premeditada, Israel lanzó un ataque directo contra infraestructura militar y nuclear en Irán, marcando así el inicio de una escalada sin precedentes en Medio Oriente. Pero lo que parecía una ofensiva unilateral y fulminante, pronto viró en dirección contraria. En cuestión de horas, Irán contraatacó con una contundencia inesperada, alcanzando objetivos estratégicos en territorio israelí. Desde ese día, las sirenas en Israel no dejaron de sonar, convirtiéndose en el eco constante de una guerra que desbordó todos los pronósticos. Por primera vez en años, el equilibrio regional parecía inclinarse fuera del mandato occidental.

Sin embargo, mientras se desplegaban misiles en el cielo, otra, aunque pequeña batalla se libraba en los entornos digitales: la de los discursos, en plataformas como YouTube donde usuarios de todo el mundo expresaron sus posturas, miedos y lealtades. Este artículo analiza más de 120,000 comentarios recogidos durante el conflicto, utilizando técnicas de procesamiento de lenguaje natural y aprendizaje automático (BERT). La intención no es solo cuantificar opiniones, sino comprender las lógicas ideológicas y algoritmos que las amplifican o silencian.

Lo que emerge de este análisis es revelador: Irán, tradicionalmente invisibilizado en los medios globales, irrumpe en la escena digital como actor discursivo central, aunque bajo una narrativa frecuentemente criminalizante. A su vez, Israel mantiene un tratamiento favorable, incluso cuando los comentarios críticos aumentan. Este trabajo plantea que el sesgo mediático y algorítmico no es una excepción, sino parte estructural del modo en que se construye la realidad en las plataformas digitales dominadas por intereses occidentales.

# REVISION DE LA LITERATURA

Los estudios sobre análisis de discurso digital en contextos de conflicto han crecido de forma significativa en la última década, particularmente en plataformas como Twitter y YouTube. Investigaciones como las de Lazer et al. (2018) y Howard y Woolley (2016) han evidenciado cómo las narrativas en línea son moldeadas por algoritmos que priorizan contenidos polarizantes o afines a intereses hegemónicos, facilitando la difusión de desinformación. Para complementar lo antedicho, Papacharissi (2015) ha planteado que las plataformas digitales no son espacios neutrales, sino estructuras afectivas que dan forma a la opinión pública según marcos culturales y políticos dominantes. En escenarios como el conflicto de Medio Oriente, esto implica que ciertos actores (como Irán) podrían ser abordados históricamente invisibilizados o representados de forma estigmatizante, mientras que otros (como Israel o EE.UU.) podrían ser presentados como protagonistas legítimos del debate global. Entonces para dar fe o partir de manera dialéctica de las tesis expuestas, se adoptan metodologías mixtas para estudiar estos fenómenos, que combinan análisis cualitativo del discurso con técnicas cuantitativas como el modelado de tópicos o la clasificación automatizada por PLN.

**LA FILOSOFIA Y LA CIENCIA DE DATOS**

El análisis de discursos sobre conflictos geopolíticos en plataformas digitales como YouTube no puede abordarse únicamente desde una lógica técnica de clasificación y visualización de datos. Es necesario incorporar una mirada filosófica crítica que permita problematizar las condiciones estructurales, ideológicas y materiales que subyacen a la producción de dichos discursos. En este sentido, la articulación entre la ciencia de datos y el materialismo dialéctico ofrece un marco robusto para comprender cómo los datos no solo reflejan la realidad, sino que también la configuran y reproducen en función de intereses sociales e históricos.

Desde la ciencia de datos, el uso de modelos de procesamiento del lenguaje natural como BERT y XLM-RoBERTa permite capturar patrones, tendencias y polaridades en los discursos emitidos por los usuarios, facilitando su clasificación automatizada y el análisis de su evolución en el tiempo. Sin embargo, como advierte Boyd & Crawford (2012), los datos nunca son neutros: están anclados en contextos de producción, mediación y poder. La elección de qué comentarios analizar, qué categorías construir y cómo interpretarlas está atravesada por supuestos epistemológicos y políticos.

Aquí es donde la concepción filosófica adquiere un papel crucial para interpretar los discursos en plataformas digitales. Una correcta perspectiva filosófica nos permite entender que los enunciados que circulan en redes como YouTube no son simples expresiones libres, sino manifestaciones concretas de una lucha de clases trasladada al campo simbólico y mediático. Aníbal Ponce (1935), por ejemplo, ya señalaba que la ideología no es una simple representación de ideas, sino una herramienta de clase que se manifiesta incluso en los productos culturales más cotidianos, configurando percepciones sociales y políticas. Desde este marco, los discursos digitales en torno al conflicto Irán–Israel deben ser leídos como expresiones de un conflicto más amplio entre proyectos civilizatorios: uno basado en la autodeterminación y resistencia de los pueblos del Sur global, y otro sostenido por la hegemonía del capital imperialista.

Incluso Žižek (2011) aunque alejado en gran parte del pensamiento marxista, aporta también una lectura complementaria al resaltar que los sujetos contemporáneos participan de estructuras simbólicas que "les hablan" antes de que ellos puedan hablar: las narrativas hegemónicas se presentan como sentido común, naturalizando lo ideológicamente construido. Así, las plataformas como YouTube no solo son espacios de expresión, sino de interpelación ideológica, donde el usuario internaliza una narrativa sobre "el enemigo" —como Irán— antes incluso de desarrollar una postura crítica propia.

Por su parte, Anwar Shaikh (2016), desde la economía política clásica, ha demostrado cómo la dinámica del capital global impone una racionalidad de eficiencia y competencia que se traduce en lo simbólico como una lucha por la legitimidad. Irán, al no alinearse con la lógica del mercado liberal global, es representado como irracional o incluso "terrorista", construyendo así una imagen funcional al relato económico-político del bloque occidental. Esto se traduce en los discursos digitales que, al reproducir la visión de medios alineados con intereses geoestratégicos, refuerzan una jerarquía simbólica entre los países del Norte y del Sur.

## METODOLOGÍA

El estudio adopta un enfoque metodológico mixto con diseño de triangulación, lo cual permite integrar de forma simultánea procedimientos cuantitativos y cualitativos (Creswell y Creswell, 2017). La parte cuantitativa se implementó mediante técnicas de procesamiento del lenguaje natural (PLN), mientras que el componente cualitativo se fundamentó en un análisis crítico y contextual de los comentarios, lo cual contribuyó tanto a la etiquetación manual como al entrenamiento supervisado del modelo. Además, se incorporó un análisis de las agendas mediáticas, siendo el contexto de conflictos geopolíticos como el de Irán e Israel, los medios no solo seleccionan qué temas visibilizar, sino que moldean las narrativas para favorecer los intereses económicos, políticos y estratégicos de las potencias imperialistas, justificando intervenciones y encubriendo prácticas coloniales (Lenin, 1916). Este enfoque permite entender cómo la representación mediática del conflicto no es neutral, sino funcional a la reproducción de relaciones de poder a nivel global.

Con el fin de operacionalizar estas ideas, y siguiendo los lineamientos metodológicos planteados por Venturini (2010) y Venturini y Munk (2022), esta investigación desarrolló su metodología bajo los siguientes procedimientos:

En primer lugar, se realizó una recopilación sistemática de comentarios sin restricción de idiomas, extraídos de vídeos de canales de noticias en YouTube y los llamados creadores de contenido o youtubers. La selección de estos vídeos se basó en las búsquedas de términos clave como "Estados Unidos", "Palestina", "Irán" e "Israel", "acotando el periodo de análisis desde el 30 de mayo al 08 de julio del 2025, es decir se tomaron 2 semanas antes y 2 semanas después del conflicto que se considera que tuvo una duración de 12 días, siendo estos desde el 13 al 24 de junio. Esta delimitación temporal se consideró pertinente por abarcar el "antes", "durante" y "después" de la controversia y sus manifestaciones discursivas iniciales. Dada la naturaleza computacional del estudio, se garantizó que los datos recolectados estuvieran debidamente contextualizados, organizados y listos para su análisis mediante técnicas de procesamiento de lenguaje natural (PLN).

En segundo lugar, se ejecutó un proceso de preprocesamiento de la base de datos, eliminando comentarios irrelevantes, vacíos o aquellos compuestos únicamente por caracteres o emoticonos sin relación directa con el conflicto, así como aquellos publicados fuera del periodo de análisis definido.

En tercer lugar, correspondió a un proceso de corrección manual basada en caracteres, a partir de una muestra aleatoria de comentarios categorizados primeramente sin entrenamiento alguno. Esta selección aleatoria permitió minimizar sesgos por concentración temática en determinados vídeos, minimizando el porcentaje de error que ya de por si era bajo.

Finalmente, se organizaron los comentarios en categorías según su contenido discursivo, considerando además su frecuencia, distribución temporal y nivel de interacción medido a través de "me gusta".

Para la recopilación de datos, se seleccionaron 3000 vídeos publicados en YouTube solo en español, inglés y francés, empleando como criterios de búsqueda términos clave relacionados con los principales actores del conflicto: "Estados Unidos", "Palestina", "Irán" e "Israel". Esta estrategia permitió garantizar que el contenido analizado estuviese directamente vinculado a la controversia en cuestión y fuera accesible para un público hispanohablante, inglés y francófono. El conjunto de datos obtenido se compuso de 120,000 registros distribuidos en cinco variables: identificador del vídeo, texto del comentario, nombre del usuario, cantidad de interacciones (likes), fecha de publicación.

La mayoría de los vídeos proceden de medios de comunicación internacionales como también se consideraron, casi en igual medida, materiales producidos por creadores independientes y vídeos educativos destinados a una audiencia general. La descarga de los comentarios se efectuó mediante la API oficial de YouTube (Data API v3) el 8 de julio de 2025 -menos de 2 semanas de terminado el conflicto-, el que se haya hecho en esta fecha se debe al afán de prever el posible descarte de videos, que reducen la cobertura de 20 a 60 días (Pusey-Reid y Ciesielski, 2024).

La determinación de las etiquetas se realizó mediante un análisis manual, sin la asistencia de técnicas automáticas de minería de entidades, enfocándose en identificar patrones recurrentes y estructuras textuales que reflejaran diferentes posturas discursivas. Este proceso se complementó con el enfoque teórico propuesto por Rico-Sulayes (2025), quien retoma los planteamientos de van Dijk (2011) y Jiwani y Richardson (2011) para comprender cómo los discursos en situaciones de conflicto político se estructuran bajo una lógica de oposición ideológica. Según este autor, los conflictos discursivos no se reducen a posturas dicotómicas simples, sino que se articulan a través de un cuadro ideológico compuesto por estrategias discursivas diferenciadas: cada grupo busca resaltar sus virtudes, minimizar sus defectos y, de manera recíproca, exponer los defectos del otro y minimizar sus virtudes. Esta perspectiva resulta especialmente útil en el análisis de comentarios en redes sociales, ya que permite identificar múltiples dimensiones de posicionamiento ideológico, no limitadas a un eje binario a favor o en contra, sino abiertas a una clasificación más rica y no excluyente. Bajo este marco teórico, el presente estudio establece etiquetas diferenciadas que permiten captar apoyos o rechazos explícitos hacia actores específicos del conflicto —como Palestina, Israel, Irán o Estados Unidos—, evitando así una simplificación forzada de las posturas discursivas y permitiendo una representación más precisa y matizada de la polarización en redes sociales.

En este estudio se empleó la estrategia metodológica conocida como "anotación basada en caracteres" (Pustejovsky y Stubbs, 2012), la cual se centra en registrar únicamente aquellos rasgos lingüísticos que son observables de manera directa en el texto, dejando de lado interpretaciones subjetivas o dependientes del contexto. Esta decisión metodológica busca facilitar la automatización del proceso de clasificación y mejorar la capacidad del modelo para generalizar sobre nuevos datos. Además, cada categoría identificada fue asignada a un código numérico específico, lo cual permite su procesamiento eficiente mediante algoritmos de aprendizaje automático.

Enseguida, exponemos cada etiqueta:

**Comentarios no relacionados (NR – Etiqueta: 0):** Esta categoría comprende aquellos comentarios que no guardan vínculo con la controversia central analizada. Incluye mensajes extraídos de vídeos en los cuales, además del conflicto principal, se tratan otros temas internacionales, locales o de entretenimiento. También se consideran aquí los comentarios que critican a periodistas, presentadores o medios sin hacer referencia directa a los actores o hechos del conflicto Hamás-Israel-Irán-EE.UU.

**Comentarios sin postura clara (SP – Etiqueta: 1):** Abarca comentarios que no manifiestan una posición explícita hacia ninguno de los actores involucrados en la controversia. Son mensajes que expresan neutralidad o promueven la paz en términos generales, por ejemplo: "Paz para todos" o "No a la guerra". También se incluyen aquellos que responsabilizan indistintamente a todas las partes del conflicto o desvían la discusión hacia temas ajenos, sin tomar postura definida hacia Israel, Palestina, Irán o Estados Unidos.

**Comentarios a favor de Israel (Pro-Israel – Etiqueta: 2):** Esta etiqueta reúne comentarios que expresan apoyo hacia el Estado de Israel. Son mensajes que justifican sus acciones militares, reivindican su derecho a la defensa o destacan su legitimidad como Estado. Ejemplos típicos incluyen frases como "Israel tiene derecho a defenderse" o "Dios bendiga a Israel", así como comentarios que denuncian ataques en su contra por parte de Hamás, Palestina o Irán.

**Comentarios a favor de Estados Unidos (Pro-EE.UU. – Etiqueta: 3):** Incluye los comentarios que defienden la participación, apoyo o política de Estados Unidos dentro del conflicto. Esta categoría recoge mensajes que muestran simpatía hacia el rol estadounidense en el conflicto, resaltan su influencia positiva o destacan su intervención como necesaria. También se incluyen expresiones de afinidad cultural, política o militar hacia Estados Unidos.

**Comentarios a favor de Palestina o Hamás (Pro-Palestina – Etiqueta: 4):** Esta categoría contiene comentarios de respaldo al pueblo palestino o al grupo Hamás, entendiendo que para ciertos discursos apoyar a Hamás se percibe como defensa de la causa palestina. Son mensajes que denuncian ataques contra Gaza, reclaman justicia para Palestina o apelan a la resistencia contra la ocupación. Se incluyen expresiones como "Palestina libre" o "Resistencia legítima", así como elogios directos a Hamás.

**Comentarios a favor de Irán (Pro-Irán – Etiqueta: 5):** Se engloban aquí comentarios que respaldan a Irán, ya sea en términos políticos, culturales o militares. Los mensajes suelen destacar el papel de Irán como opositor al sionismo o al intervencionismo occidental, así como valoraciones positivas sobre su liderazgo y resistencia frente a Estados Unidos e Israel.

**Comentarios críticos hacia Israel (Anti-Israel – Etiqueta: 6):** Esta categoría agrupa los comentarios que expresan una postura negativa hacia Israel, considerándolo responsable directo de crímenes o abusos. Son frecuentes las denuncias de "genocidio", "colonialismo" o "opresión", así como acusaciones contra su gobierno, líderes políticos o militares por las acciones realizadas en el conflicto.

**Comentarios críticos hacia Estados Unidos (Anti-EE.UU. – Etiqueta: 7):** Incluye mensajes que responsabilizan a Estados Unidos de promover guerras, intervenir en Medio Oriente o actuar en complicidad con Israel. Estos comentarios suelen manifestar descontento hacia su política exterior, sus intervenciones militares o su influencia en el desarrollo del conflicto.

**Comentarios críticos hacia Palestina o su población (Anti-Palestina – Etiqueta: 8):** Abarca comentarios que responsabilizan a la población palestina o a sus líderes por los actos violentos ocurridos. Incluye discursos que califican de terroristas a los palestinos, minimizan sus reclamos humanitarios o los culpan directamente por la escalada de violencia. Se pueden encontrar generalizaciones negativas hacia su cultura, religión o política.

**Comentarios críticos hacia Irán (Anti-Irán – Etiqueta: 9):** Contempla los comentarios que acusan a Irán de ser causante o agravante del conflicto, lo señalan como patrocinador del terrorismo o critican su gobierno y política exterior. Se incluyen mensajes que responsabilizan a Irán por el aumento de la tensión regional, así como aquellos que lo desacreditan por sus posiciones ideológicas o religiosas.

**Entrenamiento supervisado multilingüe con XLM-RoBERTa**

Para complementar la clasificación realizada con la API de OpenAI, se diseñó una etapa de entrenamiento supervisado empleando modelos BERT multilingües, a fin de evaluar la coherencia categorial en diferentes contextos idiomáticos (inglés, español y francés). Se utilizó como punto de partida el modelo XLM-RoBERTa base (Conneau et al., 2020), adaptado posteriormente mediante fine-tuning sobre una muestra estratificada de 1,800 comentarios extraídos de los datos recolectados entre el 30 de mayo y el 8 de julio de 2025. Las etiquetas empleadas en el entrenamiento fueron las mismas cinco categorías discursivas del modelo inicial, pero codificadas numéricamente para su procesamiento automatizado.

El entrenamiento se realizó en Google Colab Pro con aceleración por GPU Tesla T4, durante 6 épocas, con una tasa de aprendizaje de 3e-5 y tamaño de lote de 16. El conjunto fue dividido en 70% para entrenamiento y 30% para validación, aplicando técnicas de submuestreo para nivelar la frecuencia de las categorías menos representadas. El modelo alcanzó una precisión macro promedio de 82% y un F1-score global de 79%, evidenciando buena capacidad de generalización. Sin embargo, al igual que en experiencias previas con modelos cerrados, se identificó una alta tasa de confusión entre las categorías de "justificación social" y "normalización o resignación", especialmente en comentarios con un tono ambiguo o resignado que podía interpretarse de manera dual.

La aplicación del modelo XLM-RoBERTa permitió contrastar los resultados con los obtenidos previamente mediante la clasificación con OpenAI, revelando áreas de convergencia, pero también diferencias relevantes en el tratamiento de comentarios multilingües. Esta estrategia fortaleció la triangulación computacional, validando la robustez de las categorías y permitiendo afinar futuras tareas de anotación automática en contextos multilingües y geopolíticamente sensibles.

## Contexto del conflicto

**Origen del conflicto**

El conflicto actual entre Irán e Israel encuentra sus raíces en acontecimientos históricos que se remontan a inicios del siglo XX. Durante la Primera Guerra Mundial, entre 1917 y 1919, entre 6 y 8 millones de iraníes murieron de hambre, a pesar de que Irán se mantenía neutral. El ejército británico fue directamente responsable de esta devastación, lo que dejó una huella profunda en la memoria colectiva iraní (Majd, 2013; Abbasi, 2015).

Posteriormente, el Reino Unido instauró a Reza Chah como monarca, gobernando entre 1925 y 1941, y luego impuso a su hijo Mohammad Reza Pahlevi. En 1951, el primer ministro Mohammad Mossadeq nacionalizó el petróleo iraní, acción que provocó la intervención conjunta del Reino Unido y Estados Unidos mediante la "Operación Ajax", que culminó con el derrocamiento de Mossadeq en 1953 (Roosevelt, 1979; Gasiorowski et Byrne, 2004; Ebrahimi, 2016).

Tras el golpe, el general Fazlollah Zahedi fue colocado en el poder, y con apoyo del Mosad, se creó la temida policía política SAVAK, con asesoría israelí, lo cual cimentó la percepción negativa de Israel entre los iraníes (Sale, 1977; Irnberger, 1978).

Ya en el periodo de la Guerra Fría, Estados Unidos, preocupado por las aspiraciones expansionistas israelíes, promovió una alianza entre Irán y Siria en 1953 para contrarrestar la influencia sionista en la región (Lesch, 1992). Sin embargo, tras la revolución iraní de 1979, impulsada por la administración Carter mediante el regreso de Ruhollah Khomeini desde Francia, Irán adoptó una postura anticolonial y antioccidental (Gil Guerrero, 2016).

Durante la guerra Irán-Irak en los años 80, Israel inicialmente apoyó a Irán con armamento, en el marco del escándalo Irán-Contras (Bergman, 2008). No obstante, con el tiempo Tel Aviv cambió de estrategia. A pesar de ello, subsistieron vínculos económicos como la copropiedad del oleoducto Eilat-Ascalón, cuya información se mantiene clasificada por ley en Israel desde 2018 (Red Voltaire, 2018).

Tras la invasión de Irak por Estados Unidos y el Reino Unido en 2003, se intentó vincular a Irán con los atentados del 11 de septiembre y con un supuesto programa nuclear militar, lo cual fue desmentido posteriormente. A pesar de ello, estas acusaciones llevaron a sanciones internacionales y resoluciones del Consejo de Seguridad de la ONU en 2006 y 2007 (Sokolski et Clawson, 2004; Delpech, 2009; Patrikarakos, 2012).

En 2013, Estados Unidos e Irán negociaron el Plan de Acción Integral Conjunto (JCPoA), con la mediación de William Burns y el respaldo de Barack Obama. El acuerdo no buscaba restablecer relaciones diplomáticas sino contener un potencial peligro nuclear (Goldberg, 2016; Daalder, Gnesotto & Gordon, 2006; Parsi, 2017).

Sin embargo, la posterior retirada unilateral del acuerdo por parte del presidente Donald Trump reactivó las tensiones. Irán, en respuesta, comenzó a fortalecer alianzas con actores no estatales de la región, aunque sin control directo, en línea con la doctrina de Khomeini (Fayazmanesh, 2008; Murray, 2009).

El 7 de junio de 2025, Irán declaró haber accedido a documentos confidenciales sobre el programa nuclear israelí, revelando supuesta colaboración ilícita del director del OIEA, Rafael Grossi, con Israel (Netanyahu, 2018). El 12 de junio, el OIEA declaró que no podía garantizar que el programa nuclear iraní fuera completamente pacífico, lo que llevó la cuestión ante el Consejo de Seguridad (OIEA, 2025).

La reacción israelí fue inmediata: inició una ofensiva militar contra Irán al día siguiente, 13 de junio. Esto recuerda la estrategia israelí de 2006 contra el Líbano, utilizada entonces para frenar investigaciones sobre redes de espionaje (Meyssan, 2007). Luego siendo o no un factor decisivo el de frenar dichas investigaciones, puede haber sido solamente un detonante a los siempre vigentes planes de Israel y Estados Unidos de suplantar la facción iraní en el poder, por otra facción más amigable o servil al país hegemónico en la actualidad.

**Cronología diaria del conflicto**

**13 de junio de 2025**

Israel lanza la Operación León Naciente, una serie de cinco oleadas de bombardeos aéreos con más de 200 aeronaves y 330 municiones contra unas 100 instalaciones militares y nucleares iraníes (sitios como Natanz, Fordow e Isfahán). Fueron asesinados altos mandos militares y científicos nucleares (Saab y White, 2025).

El Mossad habría sabotado defensas aéreas e infraestructura de misiles, utilizando bases de drones encubiertas y misiles de precisión, asegurando superioridad aérea (Saab & White, 2025).

En respuesta, Irán lanza drones y misiles balísticos contra ciudades israelíes, siendo interceptados durante gran parte, aunque causaron daños menores (Al Jazeera, 2025).

**14–15 de junio**

Los combates continúan: desde Irán se lanzan oleadas de misiles y drones; Israel intercepta la mayoría mientras alerta a civiles sobre la posibilidad de evacuaciones (Al Jazeera, 2025).

**16 de junio**

Israel intensifica la ofensiva con ataques a convoyes de armas, hospitales y medios estatales iraníes; al menos ocho muertes y cientos de heridos reportados en Haifa y otras ciudades israelíes (AP News, 2025).

Hay un intento de asesinato contra el presidente iraní Masoud Pezeshkian durante una reunión del Consejo Supremo de Seguridad, quien resulta levemente herido (Reuters, 2025).

Irán responde atacando una refinería en Haifa y ejecutando a un presunto espía del Mossad (AP News, 2025).

**17–18 de junio**

Israel anuncia evacuaciones en Teherán; continúan ataques aéreos contra objetivos estratégicos iraníes (Saab & White, 2025).

Irán confirma haber derribado al menos un dron israelí en su territorio nocturno (Saab & White, 2025).

**19 de junio**

Según el informe de Amnistía Internacional (2025) El 19 de junio de 2025, Irán respondió a los ataques masivos israelíes con armamento de alto impacto, presuntamente incluyendo municiones de racimo, en zonas previamente utilizadas con fines militares. Aunque Amnistía Internacional criticó el uso de este armamento, Irán sostuvo que sus acciones fueron una defensa legítima frente a la agresión y que se dirigieron a infraestructuras estratégicas, no a civiles.

**22 de junio**

Estados Unidos entra en la guerra con la Operación Martillo de Medianoche, lanzando ataques coordinados con aviones B-2 Spirit y misiles Tomahawk contra tres sitios nucleares iraníes: Natanz, Fordow e Isfahán. Se emplearon bombas perforantes de hasta 30,000 lb, causando daños severos, mientras Irán contraatacó bombardeando la base estadounidense en Catar (Washington Post, 2025).

**23 de junio**

Irán realiza un ataque misilístico contra la base Al Udeid en Catar, considerada una represalia simbólica ante las acciones de EE. UU. (Al Jazeera, 2025).

**24 de junio de 2025**

Mediada por Estados Unidos y Catar, se establece un **cese al fuego escalonado** que inicia alrededor de las 04:00 GMT: Irán cesa primero, seguido por Israel tras unas 12 horas (Al Jazeera, 2025).

Finaliza el conflicto tras 12 días de intensa hostilidad ( Washington Post, 2025).

**Saldos del conflicto**

Las cifras oficiales informan al menos 935 muertos en Irán durante los 12 días de combates, incluyendo 132 mujeres y 38 niños, aunque organizaciones independientes como Human Rights Activists, estiman hasta 1 190 víctimas fatales y 4 475 heridos (HRANA, 2025).

El desglose sugiere que una parte significativa fueron civiles, y grupos de defensa confirmaron que murieron al menos 182 civiles, entre ellos 68 mujeres y 45 niños

En cuanto a Israel, fuentes sionistas como el Times of Israel (2025) reportan 28 a 29 muertes confirmadas (28 civiles y 1 militar fuera de servicio), con más de 3 238 personas hospitalizadas por heridas. Pero existen muchas sospechas de parte de otros medios conocidos, como HispanTV, Telesur y Russia Today, de que el gobierno israelí ha minimizado estas cifras.

**Una visión del conflicto**

En el contexto de una creciente tensión regional, el ex diplomático británico Alastair Crooke reveló que los ataques cibernéticos y de drones lanzados contra Irán fueron planificados durante años, en una operación compleja que incluyó la colaboración de Azerbaiyán, Erbil (Kurdistán iraquí), el grupo MEK y tecnologías estadounidenses, incluyendo software satelital y terrestre (Crooke, 2025). El ataque se produjo el día posterior a una supuesta acción encubierta del director del OIEA, Rafael Grossi, acusado por Irán y Rusia de filtrar información estratégica al Mossad.

Los ciberataques paralizaron la defensa iraní durante ocho horas, afectando severamente sus sistemas, aunque Irán logró reactivarse y contraatacar. Las represalias, según Crooke, incluyeron daños ocultos por Israel, como el ataque al centro del Mossad en Herzliya, la destrucción del Instituto Weizmann y una refinería en Haifa. Además, Irán habría utilizado misiles hipersónicos, mientras que Israel y EE. UU. fallaron en su defensa al disparar 93 misiles THAAD ineficaces, con un coste estimado de 1.200 millones de dólares (Crooke, 2025).

Entre los efectos del ataque israelí, se incluyen los asesinatos selectivos de 30 altos mandos militares iraníes y 11 científicos nucleares junto a sus familias, gracias a información proporcionada por el OIEA, según denuncias. El programa de espionaje utilizado habría sido MOSAIC, desarrollado por la firma Palantir (Jalife-Rahme, 2025).

Según Crooke, el supuesto objetivo de eliminar el programa nuclear civil iraní fracasó; apenas se habría retrasado unas semanas o meses. Esta postura es compartida por el físico del MIT, Ted Postol, y por los ex agentes estadounidenses Larry Johnson y Scott Ritter. Además, se sostiene la hipótesis de que EE. UU. advirtió anticipadamente a Irán del bombardeo a sus plantas nucleares mediante canales diplomáticos indirectos, posiblemente vía Omán o la embajada suiza en Teherán (Crooke, 2025).

Finalmente, según el propio Crooke, el cese al fuego fue solicitado por Israel a través de mediación omaní, lo que suscitó críticas internas en Irán, ya que se consideraba que Israel estaba en una posición debilitada en ese momento (Crooke, 2025).

# Resultados

Descrito el contexto en el que se enmarca la controversia, nos permitimos analizar críticamente los resultados de clasificación a partir de inteligencia artificial, como se muestra a continuación:

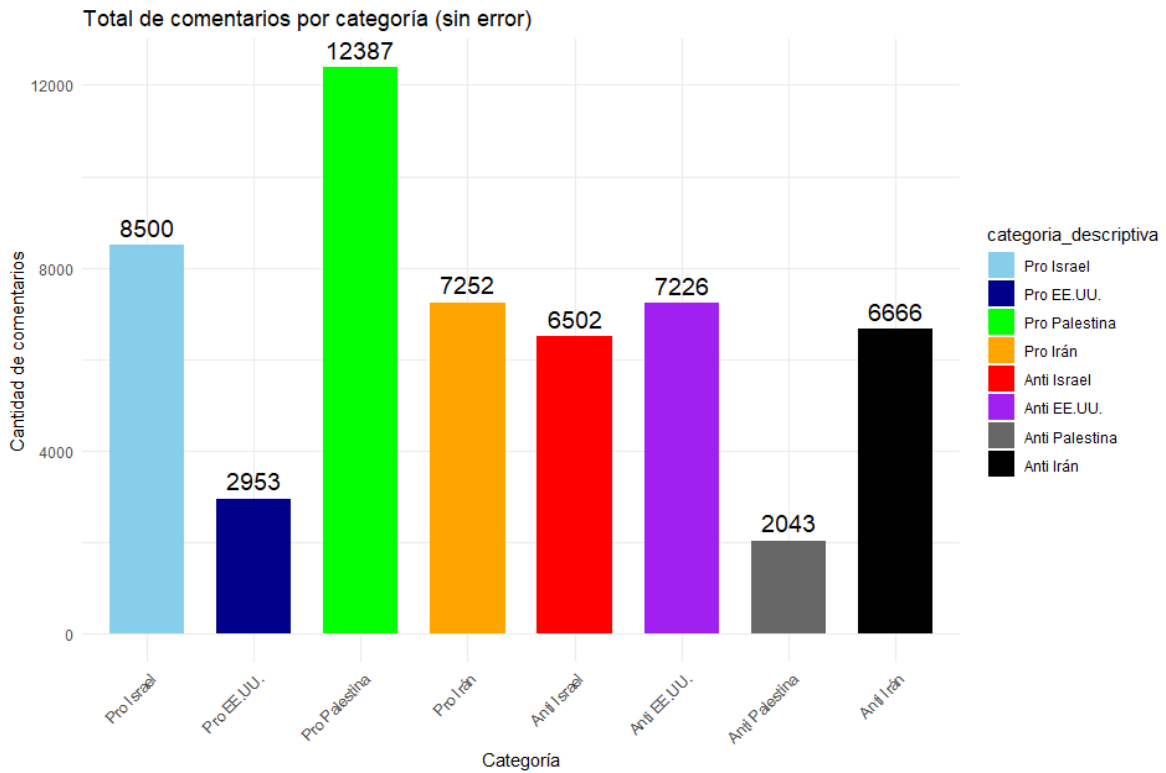

*Gráfico 1. Gráfico de barras del total de comentarios por categoría*

El gráfico evidencia una clara predominancia del discurso pro palestino, con más de doce mil comentarios, lo que sugiere una fuerte identificación del público analizado con la causa palestina frente al conflicto reciente. Esta inclinación se acompaña de un notable volumen de mensajes críticos hacia Estados Unidos e Israel, lo que refleja un clima de rechazo hacia las potencias consideradas responsables o intervencionistas. Por otro lado, aunque Irán también genera apoyos considerables, el número de comentarios en su contra es alto, revelando una polarización discursiva marcada. En contraste, el bajo nivel de respaldo a EE.UU. y la escasa presencia de críticas hacia Palestina refuerzan la idea de una narrativa dominante que ve a estos actores desde lentes muy distintas. En conjunto, el panorama discursivo sugiere que, al menos en el universo analizado, las simpatías están más alineadas con causas anti hegemónicas y narrativas de resistencia.

Said (1978) ha señalado cómo los discursos sobre Oriente Medio están impregnados de orientalismo, una construcción ideológica que representa a los países no alineados con Occidente —como Irán— como irracionales, violentos o atrasados. Esta lógica ha llevado a calificar a Irán como "terrorista" no por evidencias objetivas, sino por su voluntad de autonomía estratégica frente al orden internacional dominado por Estados Unidos y sus aliados. En este marco, la independencia política y cultural de Irán es narrada como amenaza, cuando en realidad constituye una afirmación de soberanía. Said advierte que estas representaciones no solo desinforman, sino que sirven para justificar intervenciones y sanciones bajo la apariencia de defensa de la democracia.

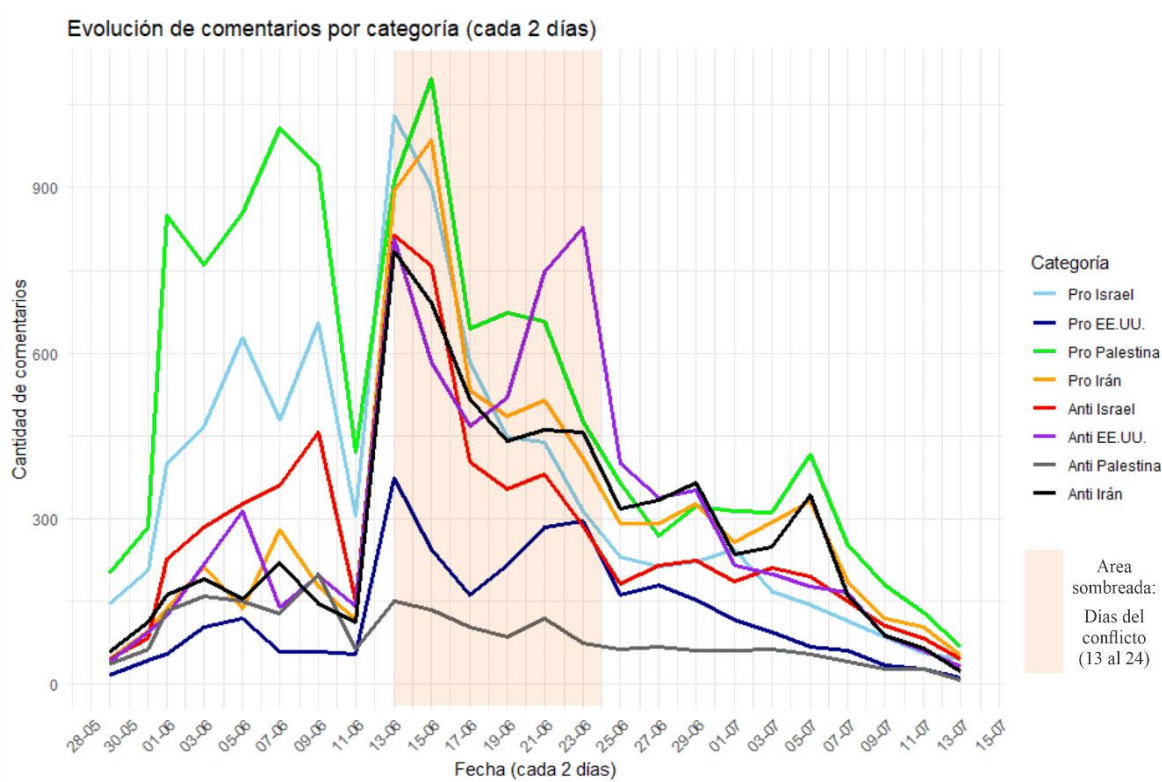

*Gráfico 2. Evolución de la cantidad de comentarios en el tiempo (cada 2 días)*

Para este gráfico, empezaremos por señalar que el descenso abrupto del día 13 de junio, es decir una vez que estalla el conflicto, no responde a un problema metodológico relacionado con sesgo de muestreo ni a limitaciones en la extracción de datos. La captura de comentarios se realizó sobre una base amplia y constante de videos vinculados al conflicto, sin modificar los criterios de selección o la cantidad de videos considerados. Por el contrario, este descenso puede explicarse por dos fenómenos característicos de contextos de alta tensión geopolítica. Primero, plataformas como YouTube tienden a modificar los parámetros de interacción, limitando o desactivando secciones de comentarios en ciertos canales o transmisiones oficiales durante eventos sensibles, lo cual reduce el volumen total visible. Segundo, estudios sobre comportamiento digital durante crisis señalan un incremento del "consumo pasivo", donde los usuarios prefieren observar contenidos sin interactuar activamente en los espacios de comentarios, especialmente durante momentos de alta incertidumbre o violencia (Liu et al., 2021). Estos factores explican por qué el volumen de comentarios disminuye por unas varias horas durante el estallido del conflicto, aun cuando la atención mediática y el consumo de contenido se mantienen o incluso aumentan.

Antes del estallido del conflicto, Irán era prácticamente ausente en la conversación pública global en plataformas como YouTube, a diferencia de actores como Israel o Estados Unidos, que gozan de una visibilidad constante debido a su protagonismo histórico en el relato geopolítico occidental. Esta invisibilización de Irán se inscribe dentro de un patrón más amplio de silenciamiento estructural hacia los países del sur global, como lo ha señalado Said (1978), quien argumenta que los discursos dominantes construyen una imagen de Oriente no solo como "otro" cultural, sino también como sujeto pasivo sin agencia histórica.

Sin embargo, con el inicio del conflicto, Irán irrumpe abruptamente en la escena discursiva digital, evidenciando un fuerte aumento tanto en expresiones de apoyo como de oposición. Este contraste sugiere que su ausencia previa no respondía a una falta de relevancia geopolítica, sino a un sesgo estructural en la representación mediática. Como señalan Herman y Chomsky (1988), los grandes medios actúan como sistemas de propaganda que seleccionan qué países merecen atención y cómo deben ser representados, en función de intereses geoestratégicos.

Plataformas como YouTube, lejos de ser espacios neutrales, amplifican estos marcos hegemónicos mediante algoritmos que priorizan contenidos alineados con perspectivas dominantes (Gillespie, 2018). Así, la repentina visibilidad de Irán se da bajo condiciones sesgadas: emerge no como un actor legítimo del sistema internacional, sino como el antagonista de una narrativa ya establecida. Esta asimetría discursiva, donde el mundo parece "de Occidente", revela una estructura mediática que continúa reproduciendo jerarquías coloniales de visibilidad, incluso en entornos digitales aparentemente abiertos (Couldry y Mejias, 2019).

En ese sentido, Irán se torna un tema de debate recién cuando se posiciona como contendiente activo en el conflicto. Aun así, se observa un cambio en la percepción: los comentarios muestran un aumento en el apoyo a Irán y un franco desacuerdo hacia Israel.

Este último país, que antes del conflicto mantenía niveles relativamente estables de apoyo, comienza a registrar un descenso sostenido durante la fase bélica. Esta evolución sugiere una transformación narrativa que desafía los marcos tradicionales pro-occidentales. Sin embargo, este análisis debe contrastarse con futuras estadísticas relacionadas con Israel, ya que se hipotetiza un posible sesgo algorítmico de YouTube que tiende a representar a Israel de forma predominantemente positiva, tal como advierte (Noble, 2018) en su análisis sobre algoritmos y reproducción de poder.

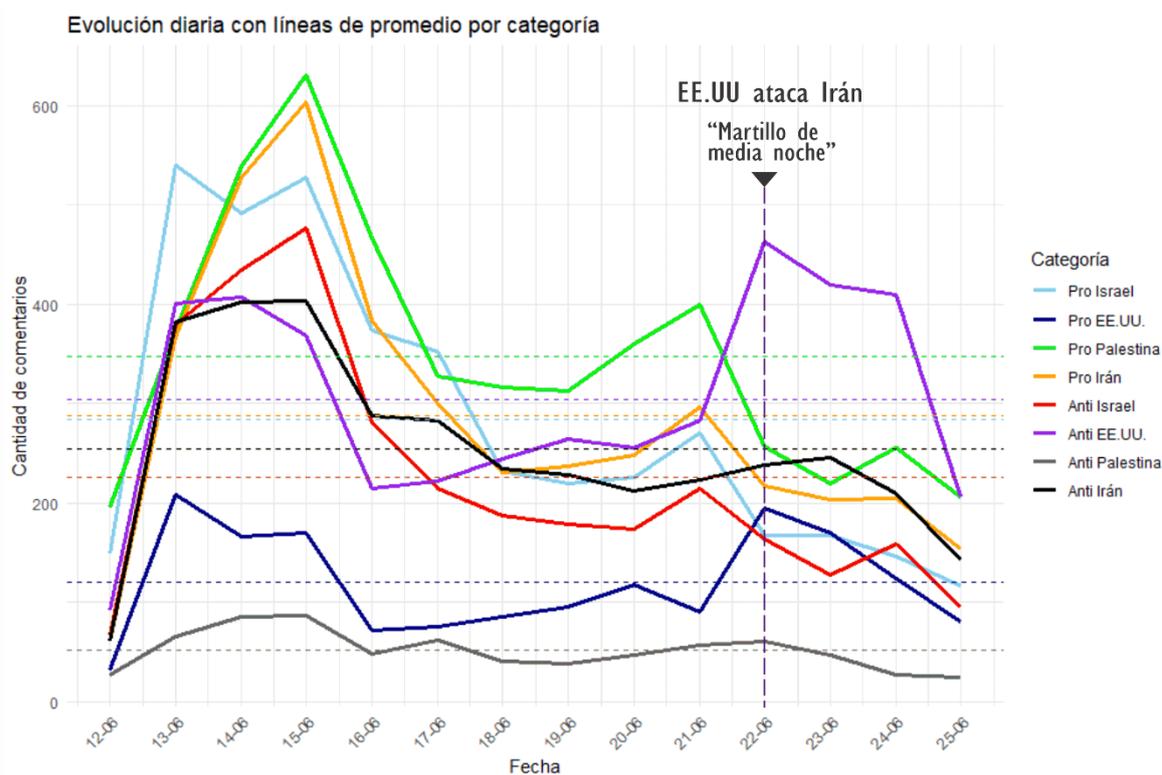

*Gráfico 3. Evolución de la cantidad de comentarios en pleno conflicto (cada día)*

Tras el inicio del conflicto (13-14 de junio), hay un repunte pronunciado en las categorías Pro Palestina y Pro Irán, superando incluso a Pro Israel, lo que evidencia una fuerte simpatía inicial hacia los actores opuestos a Israel y Estados Unidos. Esta tendencia también se acompaña de un aumento significativo en los comentarios Anti Israel y Anti EE.UU., consolidando una narrativa de crítica al eje occidental. Las categorías Pro Israel y Pro Estados Unidos, aunque altas al inicio, declinan progresivamente y no logran recuperar su nivel inicial, lo cual puede interpretarse como un debilitamiento en el respaldo popular digital conforme avanza el conflicto.

En contraste, el bloque discursivo Pro Palestina se mantiene alto durante casi todo el periodo, mostrando una resistencia simbólica constante y visibilidad sostenida, en parte alimentada por narrativas alternativas que circulan ampliamente en redes sociales.

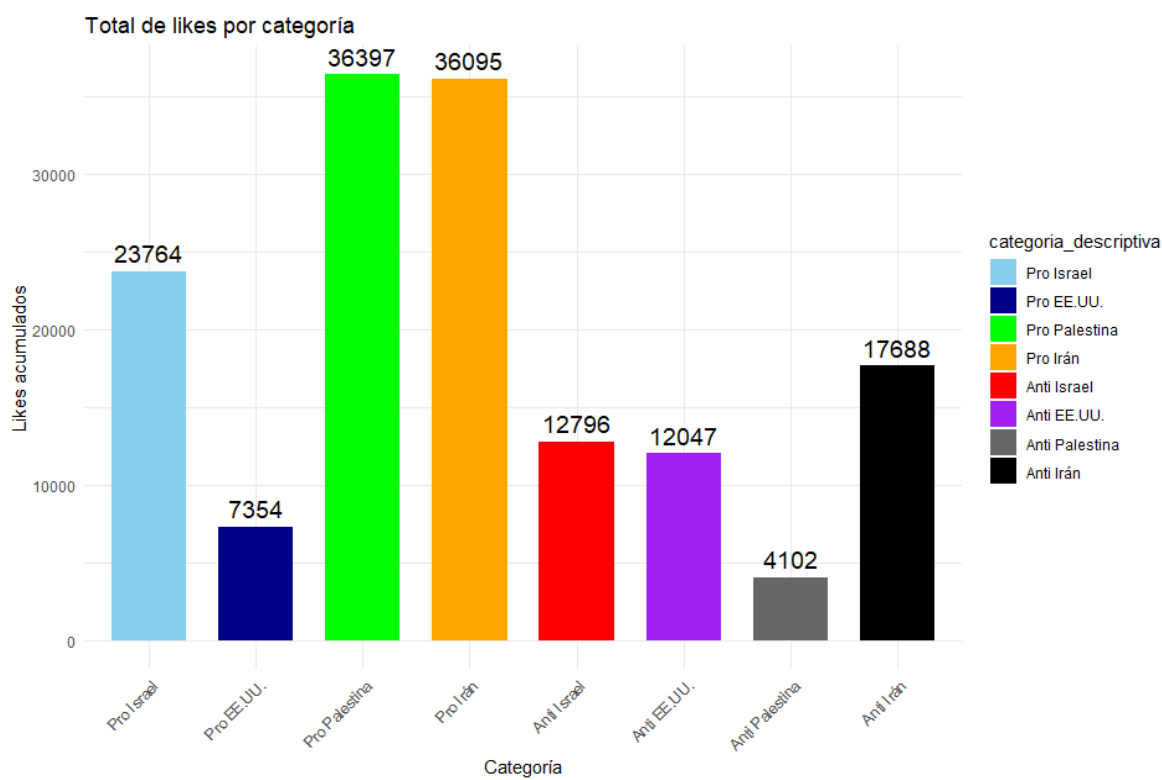

*Gráfico 4. Cantidad de interacciones (likes) por cada categoría*

El Gráfico 4 revela el nivel de apoyo explícito que reciben los comentarios de cada postura discursiva dentro del debate geopolítico digital en YouTube, medido a través de la cantidad de "like". Se evidencia una notable disparidad entre las posturas discursivas vinculadas al bloque pro-palestino y aquellas afines a Israel o Estados Unidos. A pesar de que tradicionalmente los discursos pro-israelíes han gozado de mayor visibilidad mediática y legitimación en espacios digitales, los datos revelan que los mensajes explícitamente Pro Israel y Pro EE.UU. reciben una interacción relativamente baja si se compara con el apoyo que obtienen las categorías Pro Palestina y Pro Irán. Esta diferencia no solo puede interpretarse como un cambio en la opinión pública digital, sino como una consecuencia del propio carácter del conflicto: lo que ocurre en Gaza ha sido identificado por numerosos organismos como una situación de genocidio más que un conflicto convencional, donde no hay "muertes colaterales", sino asesinatos sistemáticos y hambruna inducida.

En ese marco, el nivel de likes en las categorías Pro Israel y Pro EEUU podrían reflejar de forma directa o indirecta, estar justificando crímenes de lesa humanidad. En cambio, el apoyo mayoritario a Palestina e Irán indica una sensibilidad creciente en la audiencia digital frente a la violencia estructural ejercida sobre el pueblo palestino. Tal como afirma Žižek (2023), la ideología opera muchas veces en lo no dicho: al no declarar abiertamente el deseo de exterminio, sino camuflarlo en discursos religiosos o nacionalistas, se naturaliza la violencia. Esto resalta la necesidad de investigaciones futuras que documenten cómo estas formas de propaganda sutil construyen legitimidad discursiva en redes sociales, a pesar de la oposición creciente de una comunidad global más crítica.

En resumidas palabras, una forma de mostrarse a favor del asesinato de niños en Gaza, es comentar -durante el desarrollo del conflicto con mayor razón- vivas a Israel o EEUU. Tal vez no se puede decir esto de la totalidad de los comentarios Pro Israel y Pro EEUU, pero si consideramos que una gran parte, algo difícil de confirmar estadísticamente, pero que no es ocioso afirmar.

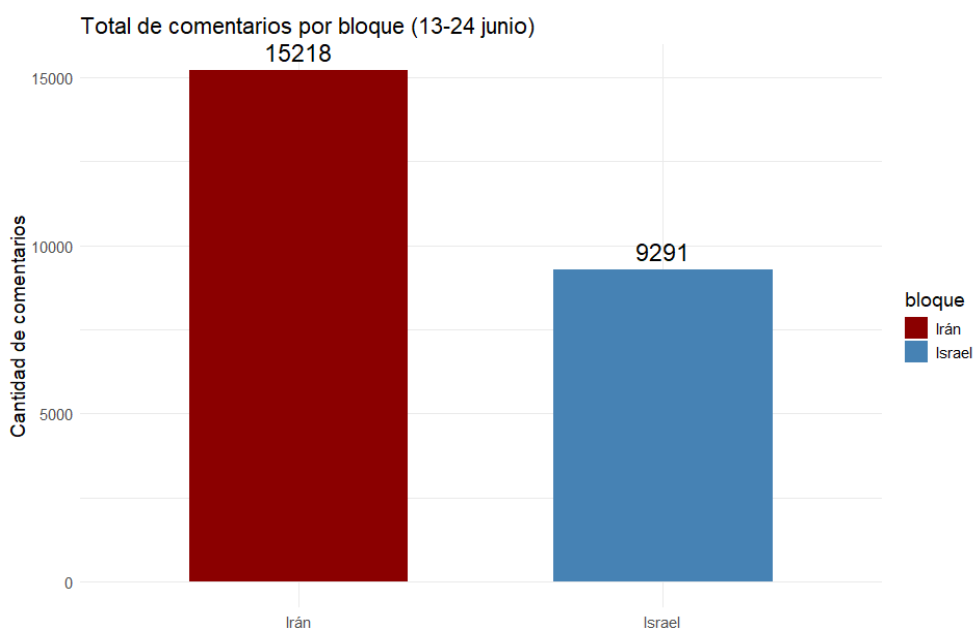

*Gráfico 5. Agrupación de categorías (Visión dicotómica)*

El gráfico expone el volumen total de comentarios durante el periodo del conflicto activo (13 al 24 de junio), diferenciando los discursos según bloques discursivos contrapuestos: Irán e Israel. Para construir estos bloques, se agruparon las categorías como siguen:

El bloque Irán está conformado por los comentarios etiquetados como Pro Irán, Pro Palestina, Anti Israel y Anti EE.UU., reflejando posturas discursivas alineadas con intereses

o simpatías hacia Irán o sus aliados geopolíticos. Por otro lado, el bloque Israel integra las categorías Pro Israel, Pro EE.UU., Anti Palestina y Anti Irán, concentrando discursos favorables a Israel y sus aliados. Esta consolidación de etiquetas permite una visualización más clara de las dinámicas de apoyo o rechazo a los dos polos del conflicto.

El gráfico muestra que durante el periodo comprendido entre el 13 y el 24 de junio, el bloque discursivo asociado a Irán acumuló 15,218 comentarios, superando ampliamente al bloque de Israel, que alcanzó 9,291 comentarios. Esta diferencia cuantitativa sugiere que, en el contexto del conflicto, Irán se posicionó como el centro principal del debate digital, a pesar de ser históricamente menos visible en plataformas occidentales. La mayor presencia discursiva del bloque Irán puede interpretarse como una reacción frente a acciones militares o narrativas mediáticas dominantes, lo que movilizó una mayor participación y visibilidad por parte de audiencias críticas al eje occidental. Asimismo, refleja que las simpatías hacia Palestina e Irán, así como las críticas a Israel y EE.UU., no solo son significativas en volumen, sino que estructuran un campo de opinión con capacidad de disputar la hegemonía narrativa del bloque proisraelí. Este gráfico permite concluir que la agenda digital no fue dominada por el discurso oficial occidental, sino que se configuró como un espacio de confrontación simbólica donde los sectores alineados con Irán y su operación "Promesa Verdadera 3", lograron mayor resonancia.

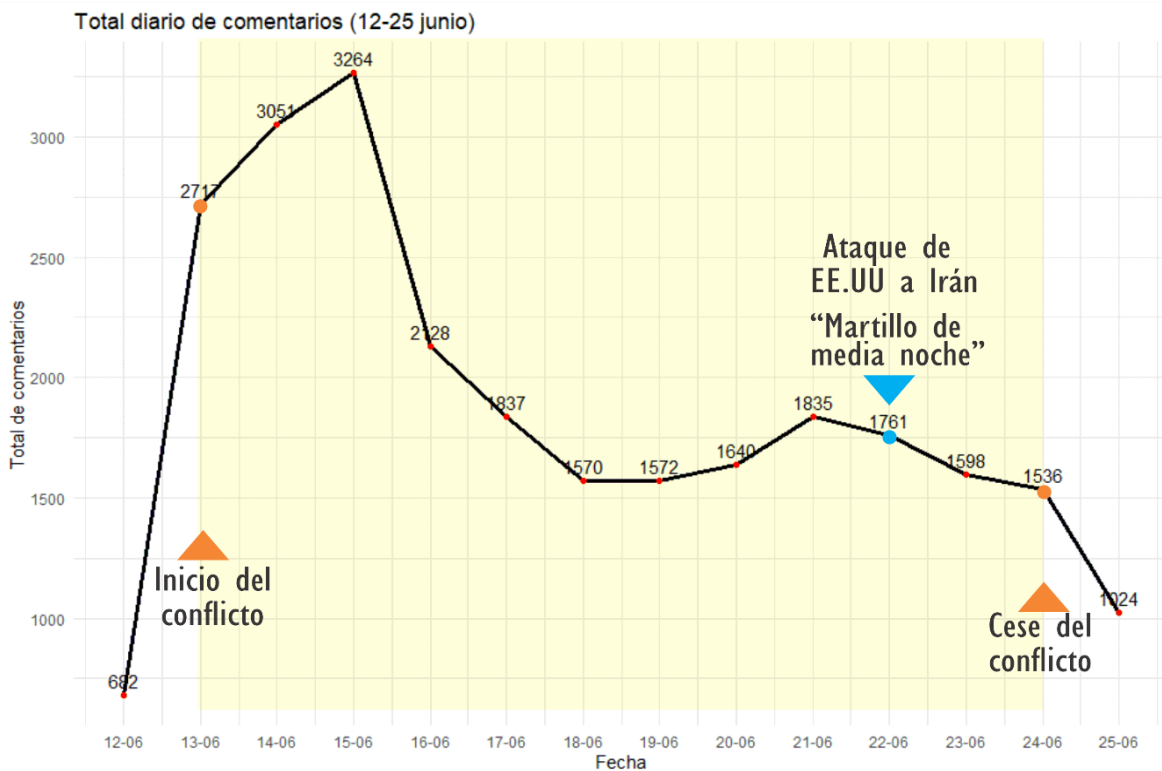

*Gráfico 6. Número total de comentarios por cada día (durante el conflicto)*

Finalmente, hacia el cese del conflicto el 24 de junio, se registra una disminución continua. Este descenso general sugiere que la intensidad discursiva en redes sociales está directamente ligada a los momentos de mayor incertidumbre y violencia, siendo el estallido inicial el que provoca la mayor efervescencia. Así, el gráfico permite concluir que la conversación digital se activa de forma inmediata ante el inicio del conflicto, pero tiende a estabilizarse e incluso a decrecer conforme se desarrollan los eventos, aun siendo algunos de dichos eventos, eventos de gravedad como el caso del ataque estadounidense a Irán, aun ello continúa inmerso en un nivel decreciente.

**Robustez estadística**

A continuación, se muestran residuos de Pearson estandarizados:

| Categoría | Observado | Esperado | Residuo de Pearson estandarizado |
|---|---|---|---|
| **Pro Palestina** | 12,387 | | **+69.63** |
| **Pro Israel** | 8,500 | | **+22.11** |
| **Pro Irán** | 7,252 | | **+6.86** |
| **Anti EE.UU.** | 7,226 | 6,691.13 | **+6.54** |
| **Anti Irán** | 6,666 | | **–0.31** |
| **Anti Israel** | 6,502 | | **–2.31** |
| **Pro EE.UU.** | 2,953 | | **–45.70** |
| **Anti Palestina** | 2,043 | | **–56.82** |

Estos permiten identificar qué categorías contribuyen significativamente a la diferencia global detectada por la prueba chi-cuadrado. Los valores superiores a +2 o inferiores a –2 se consideran estadísticamente significativos ($p < 0.05$).

- Pro Palestina y Pro Israel están fuertemente sobrerrepresentadas.
- Anti Palestina y Pro EE.UU. están fuertemente infrarrepresentadas.

Otras categorías como Anti EE.UU. y Pro Irán muestran leves pero significativas sobrerrepresentaciones. Esto confirma que existe una asimetría discursiva marcada dentro del conjunto de comentarios analizados, con predominio de posturas pro-palestinas e iraníes, y una menor presencia relativa de discursos pro-estadounidenses y anti-palestinos.

De la misma manera, cuando agrupamos 2 bloques discursivos, como se muestra en nuestro gráfico 05, tenemos:

| Bloque | Observado | Esperado | Residuo de Pearson estandarizado |
|---|---|---|---|
| **Bloque Irán** | 15,218 | 12,254.5 | +18.89 |
| **Bloque Israel** | 9,291 | | −18.89 |

Se confirma una diferencia estadísticamente significativa entre los bloques ideológicos: el bloque Irán está sobrerrepresentado y el bloque Israel, infrarrepresentado.

El residuo estandarizado (±18.89) y el valor-p < 0.001 evidencian una clara asimetría discursiva a favor del bloque Irán.

**CONCLUSIÓN**

Las dinámicas discursivas observadas en los comentarios de YouTube durante el conflicto Irán–Israel reflejan una clara transformación en la visibilidad de los actores geopolíticos. Irán, usualmente invisibilizado en la conversación global, emerge como eje central de discusión solo tras asumir un rol activo como contrincante militar. Esta irrupción discursiva no se traduce necesariamente en una visión negativa, pues las categorías de apoyo hacia Irán y sus aliados (como Palestina) crecieron considerablemente, en contraste con el decrecimiento de discursos favorables a Israel, lo que evidencia un posible giro en la percepción pública digital ante escenarios de conflicto prolongado.

El análisis cuantitativo y cualitativo de los datos evidencia además el papel determinante de los algoritmos y de las agendas mediáticas en la configuración de lo visible y lo decible. La plataforma, al amplificar o atenuar ciertos discursos, refuerza estructuras de poder donde Occidente —particularmente Israel y EE. UU.— gozan de mayor cobertura inicial y posicionamiento favorable. Sin embargo, esta hegemonía se ve desafiada por la irrupción de narrativas alternativas, que encuentran en los espacios digitales un canal de expresión, especialmente cuando los eventos bélicos rompen el orden comunicacional impuesto. Estas conclusiones refuerzan la necesidad de seguir explorando la relación entre conflictos armados, plataformas algorítmicas y opinión pública transnacional.

El análisis evidencia que, pese a la histórica hegemonía digital de Israel, el respaldo explícito es limitado; el apoyo se manifiesta de forma encubierta, mediante consignas religiosas o nacionalistas, que esquivan referirse directamente a los asesinatos y a la crisis humanitaria en Palestina.

Esta investigación utilizó PLN y un modelo supervisado entrenado con 1.400 comentarios para clasificar más de 120.000 mensajes en YouTube sobre el conflicto Irán-Israel. Al integrar herramientas computacionales con análisis contextual, se identificaron patrones ideológicos y variaciones discursivas a lo largo del tiempo, enriqueciendo la comprensión del posicionamiento digital en conflictos geopolíticos.

**SUGERENCIA A UNA FUTURA INVESTIGACIÓN**

Se recomienda que futuras investigaciones adopten un enfoque longitudinal que analice cómo YouTube ha contribuido, durante años, a reforzar una imagen positiva de Israel como país desarrollado, ocultando las estructuras de violencia y subordinación que lo sostienen. Esta narrativa se ve respaldada por potencias occidentales y también por economías del sur global. En particular, América Latina —a través del extractivismo y el trabajo precarizado— contribuye indirectamente al sostenimiento de países que subvencionan a Israel, e incluso el propio Estado israelí mantiene vínculos de explotación directa en la región, especialmente en sectores estratégicos como la vigilancia, la agricultura y la minería.

**REFERENCIAS**